\documentclass[11pt]{article}
\usepackage{amssymb}
\usepackage{graphicx}
\begin{document}

\title{\bf Competition between dust scattering albedo and 2175 \AA\ bump
for ultraviolet colours of nearby disc galaxies}
\author{Akio K. INOUE\thanks{College of General Education, 
Osaka Sangyo University, 3-1-1, Nakagaito, Daito, Osaka 574-8530, Japan;
akinoue@las.osaka-sandai.ac.jp}}
\date{}
\maketitle
\thispagestyle{empty}

\begin{abstract}
Observed ultraviolet (UV) colours of nearby disc galaxies show a
 reddening relative to their expected intrinsic colours. Since the 2175
 \AA\ bump found in the Milky Way's dust extinction law blues the UV
 colours, it might suggest that dust in nearby disc galaxies does not
 have the bump and that the Milky Way is exceptional. However, this
 conclusion can be modified by the effect of scatterings. If the
 scattering albedo decreases towards shorter wavelengths, observable UV
 colours redden. An extensive comparison between observed UV colours and
 those expected from radiative transfer simulations shows two types of
 dust suitable for nearby disc galaxies: (1) dust with a bump and a
 smaller albedo for a shorter wavelength (except for the bump range),
 and (2) dust without any bump but with an almost constant albedo. If
 very small carbonaceous grains responsible for the common unidentified
 infrared emission band are also the bump carrier, the former dust is
 favorable.
\end{abstract}

\section{Introduction}

Nearby starburst galaxies observed with the {\it IUE} satellite 
show a tight correlation between the observed ultraviolet (UV) 
spectral slope ($\beta$; $f_\lambda \propto \lambda^{-\beta}$) 
and the infrared (IR)-to-UV flux ratio (so-called IRX): 
a redder $\beta$ for a larger IRX (Calzetti et
al.~1994, Meurer et al.~1999).  Witt \& Gordon (2000)
argued that the UV colour cannot be redden by the extinction law of the
Milky Way (MW) because the strong absorption bump lies in the near-UV
(NUV). Indeed, the extinction in the NUV is slightly larger than that in
the far-UV (FUV) for the MW extinction law. Their finding may suggest 
the lack of the bump carrier in the interstellar medium (ISM) of the
starburst galaxies.

Quiescent or modest star-forming ``normal'' galaxies 
show systematically redder UV colours than those of the {\it IUE}
starburst galaxies (Bell 2002, Kong et al.~2004). This fact has recently
been confirmed by the {\it GALEX} satellite for larger samples of nearby
galaxies selected in the NUV or optical (Buat et al.~2005, Seibert et
al.~2005). According to Witt \& Gordon (2000), the MW type dust cannot
reproduce even the UV colour of the {\it IUE} starburst galaxies, much
less the redder UV colour of normal galaxies. Does the red UV colour of
normal galaxies indicate the lack of the bump carrier in their ISM and
distinguish the MW from these galaxies?

On the other hand, Granato et al.~(2000) pointed out that, even if the
bump exists in the extinction law, the strength of the bump can be
greatly reduced in the {\it attenuation law}, which is based on the
ratio of the fluxes escaped from the medium to the intrinsic ones, 
by a radiative transfer effect coupled with an {\it age-selective
obscuration}, i.e. young stars are more obscured selectively.
More recently, Panuzzo et al.~(2006) very well reproduced the red UV
colours of the {\it GALEX} galaxies with the MW type dust. They adopted
a realistic stellar distribution; younger stars are more deeply
embedded in the dust disc, whereas older stars distribute more
extensively (e.g., Robin et al.~2003). This realistic configuration
of dust and stars depending on the stellar age produces an 
{\it age-selective obscuration}. This results in a {\it steep}
attenuation law which overcomes the {\it blueing} by the bump.

In addition to the {\it age-selective obscuration}, Inoue et al.~(2006)
newly discussed the effect of the wavelength dependence of the scattering
albedo on the UV colour. In fact, the dust properties adopted by Witt \&
Gordon (2000) and Granato et al.~(2000, and also Panuzzo et al.~2006)
are different from each other, especially the wavelength dependence of
albedos. This point significantly affects the expected UV colour. 

Inoue et al.~(2006) thoroughly examined dust properties in nearby
galaxies, in particular the presence of the bump and the wavelength
dependence of the albedo, based on the {\it GALEX} colour. They adopted
a one-dimensional plane-parallel radiative transfer model developed by
Inoue (2005). While its computational geometry is one-dimensional, it
can treat the clumpiness of stars and dust thanks to the mega-grain
approximation (V{\'a}rosi \& Dwek 1999). 
Owing to the computational cheapness of the one-dimensional
calculation, they could investigate a very wide range of physical
quantities of disc galaxies. After an extensive comparison between the
{\it GALEX} data and their radiative transfer simulations, they found
two types of dust suitable for the nearby disc galaxies: 
(1) dust with a bump and a smaller albedo for a shorter wavelength
(except for the bump range), and (2) dust without any bump but with an
almost constant albedo.

This paper is a digest of Inoue et al.~(2006). If the readers need more
details, they can be found in the original paper. The next section gives
a brief summary of the radiative transfer model adopted in Inoue et
al.~(2006). A comparison of the {\it GALEX} data with the radiative
transfer simulations is presented in section 3. The last section gives a
discussion about the bump carrier.

\section{Radiative transfer model}

Inoue et al.~(2006) examined several set-ups of the star--dust
configuration in order to discuss its effect. Here we concentrate on
their standard case, the most realistic one.

A clumpy ISM is produced by a two-phase ISM model based on Wolfire et
al.~(2003); the cold and warm neutral media are regarded as clumps and
the inter-clump medium, respectively. As introduced by Inoue (2005), 
we can set the density contrast between these media and the clump
filling factor by giving the ISM mean pressure and density. The clump
size is given by its Jeans length. The gas and dust are confined in a
disc with a vertical height, and clumps distribute randomly in the
disc. Any systematic radial and vertical structures are not considered.
The ranges of physical quantities considered are the ISM mean pressure
as $10^{3.0-4.0}$ K cm$^{-3}$, the ISM mean density of hydrogen atom as
0.5--24 cm$^{-3}$, the half height of the dusty disc as 50--300 pc, and
the dust-to-gas mass ratio as 0.001--0.01. These ranges well cover the
values observed in nearby ``normal'' galaxies.

The stars are divided into three populations depending on the age: 
young (age $\leq10$ Myr), intermediate (10 Myr $<$ age $\leq300$ Myr), 
and old stars (age $>300$ Myr). The young stars are embedded in the
clumps; the emissivity of these stars is reduced by a local obscuration
factor. This is motivated by the fact that stars are formed in 
molecular clouds (i.e. clumps). The intermediate/old stars distribute
with a smaller/larger scale-height than the dusty disc height. 
This is based on the observed age-dependent stellar distribution 
(e.g., Robin et al.~2003). This age-dependent distribution produces an
age-dependent obscuration. Finally, an exponentially decaying star
formation history with 5 Gyr e-folding time and the galactic age of 10
Gyr are assumed in order to obtain the luminosity fractions of the three
stellar populations which are used as their weights in the composite
process of the transmission rates of lights from these three populations
(see equation [1] in Inoue et al.~2006).

Six dust models are compiled from the literature: two from Witt \&
Gordon (2000) (hereafter WG dusts) and four from Draine and
co-workers (hereafter Draine dusts). Table~1 is a summary
of the dust models. Witt \& Gordon (2000) empirically derived the dust
properties from a large compilation of astronomical observations. On the
other hand, Draine and co-workers made theoretical models based on
experimental optical constants of candidate materials for the
interstellar dust. Note that Draine dusts are sometimes designed to fit
astronomical observations by modifying the optical constants.
Four types of dust compositions are considered: the MW, the SMC, 
the LMC av, and LMC 2. The LMC av type is an average dust composition
over many sight lines towards the LMC, except for the supershell 
around the 30 Doradus, and the LMC 2 type is that towards the
supershell. 

Fig.~1 shows the extinction cross sections (panel [a]) and the albedos 
(panel [b]) of these models as a function of the wavelength. The
extinction cross sections (i.e. extinction laws) are very similar
between the WG dust and the Draine dust if we compare the same
composition type. We find a prominent bump at 0.22 $\mu$m 
($1/\lambda=4.6$ $\mu$m$^{-1}$) in the MW and the LMC av types, 
a weak bump in the LMC 2 type, and no bump in the SMC type. 
On the other hand, the albedos are very different between the WG dust
and the Draine dust (panel [b]). Except for the bump region, albedos of
the WG dust (solid and dashed lines) show a flat wavelength dependence in 
$2\,\mu{\rm m}^{-1}<1/\lambda<8\,\mu{\rm m}^{-1}$, whereas those of the
Draine dust (other lines) show a rapid decrease towards shorter
wavelengths. Since the albedos estimated from observations show a large
dispersion (Gordon 2004, see also Fig.~2 in Inoue et al.~2006), 
both dust models are still compatible with the data.

The radiative transfer equations are solved in a one-dimensional
plane-parallel disc geometry. The disc consists of a clumpy dusty disc
and three stellar discs described above. The clumpiness of the medium
(i.e. the dust distribution) is treated by the mega-grain approximation 
developed by V{\'a}rosi \& Dwek (1999). In this approximation, 
a dusty clump is regarded as a huge particle producing
absorption and scattering effects like a normal single dust grain. 
V{\'a}rosi \& Dwek (1999) clearly show the validity of the approximation
by comparisons between the approximate solutions and their
three-dimensional Monte Carlo radiative transfer solutions.
A set of the practical equations are presented in Inoue (2005).

\begin{table}
 \caption{Dust models.}

 \bigskip

 \centering
 \begin{tabular}{lcc}
  \hline
  Model & Reference & $k_{{\rm d},V}$ $^a$\\
  \hline
  MW (WG) & Witt \& Gordon (2000) & $2.60\times10^4$ \\
  SMC (WG) & Witt \& Gordon (2000) & $1.56\times10^4$ \\
  MW (D) & Draine (2003) & $2.60\times10^4$ \\
  SMC (D) & Weingartner \& Draine (2001) & $1.56\times10^4$ \\
  LMC av (D) & Weingartner \& Draine (2001) & $1.95\times10^4$ \\
  LMC 2 (D) & Weingartner \& Draine (2001) & $1.89\times10^4$ \\
  \hline
 \end{tabular}

 \bigskip

 $^a$ Visual extinction cross section per unit dust mass 
 (cm$^2$ g$^{-1}$).
\end{table}

\begin{figure}
 \centering
 \includegraphics[width=10cm]{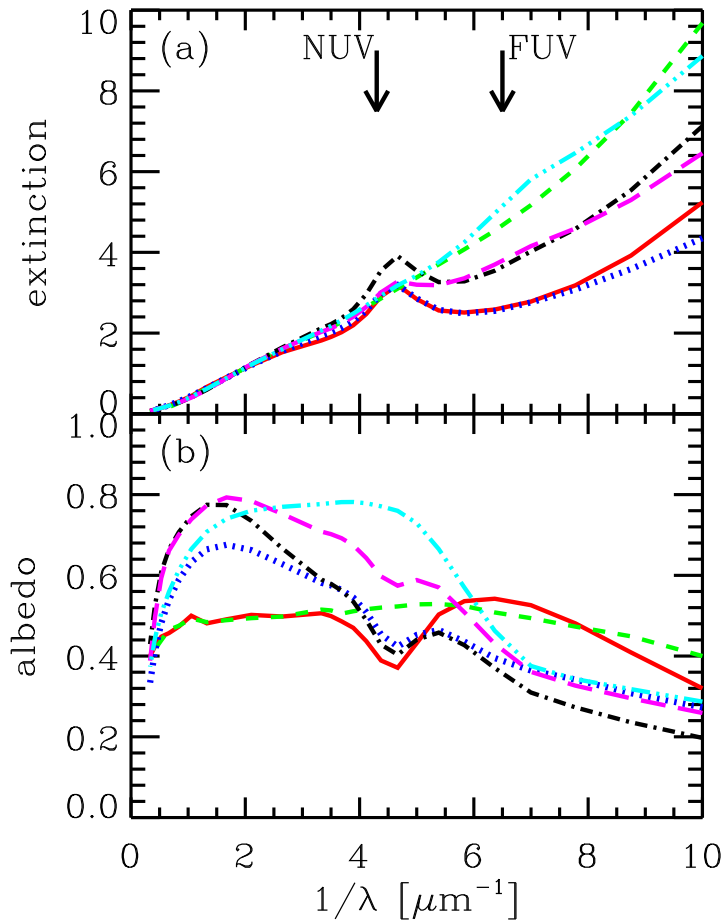}
 \caption{Differences of dust models. The panel (a) shows the extinction
 laws (extinction cross sections normalized by that at the $V$ band) 
 and the panel (b) shows the albedos. The solid and short-dashed lines
 are the MW and the SMC types of Witt \& Gordon (2000),
 respectively. The dotted, dot-dashed, long-dashed, and
 three-dots-dashed lines are the MW, the LMC av, the LMC 2, and
 the SMC types of Draine (2003) and Weingartner \& Draine (2001),
 respectively. The two downward arrows in the panel (a) show the
 effective wavelengths of the two {\it GALEX} filters.}

\end{figure}

\section{IR-to-UV flux ratio and {\it GALEX} colour}

Since the {\it GALEX} colour is very sensitive to the presence of the
bump and the wavelength dependence of the albedo, we may assess the dust
models by comparing the observed {\it GALEX} colours with those expected
from the radiative transfer simulations. 

\begin{figure}
 \centering
 \includegraphics[width=15cm]{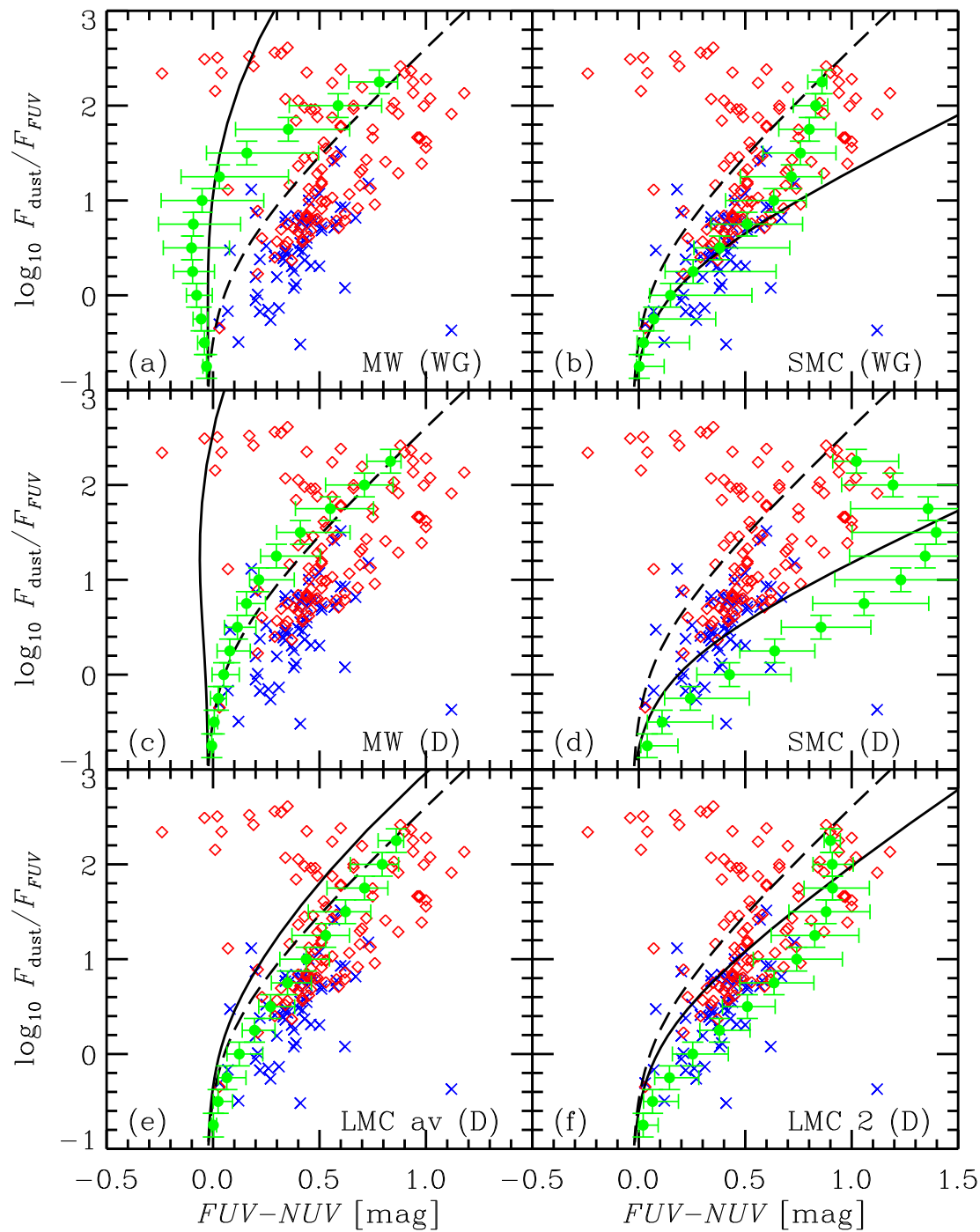}
 \caption{Dust IR-to-FUV flux ratio and the {\it GALEX} colour. 
 The crosses and diamonds are observed data of the NUV selected and the
 FIR selected nearby galaxies, respectively, taken from Buat et
 al.~(2005). The solid and dashed lines correspond to the extinction law
 and the Calzetti law, respectively. The colours predicted from the
 radiative transfer model are divided into several bins in the flux
 ratio. The vertical error-bars are the bin width, the horizontal
 error-bars are the full width of the distribution of $FUV-NUV$ in each
 bin, and the filled circles show the median location of the
 distribution.}
\end{figure}

Fig.~2 shows the diagram of the IRX (dust IR-to-UV flux ratio, $F_{\rm
dust}/F_{FUV}$) and the {\it GALEX} colour, $FUV-NUV$. The crosses and
diamonds are the observed data of the nearby galaxies selected by NUV
and FIR, respectively, taken from Buat et al.~(2005). The solid and
dashed lines are the loci expected from the extinction law and the
Calzetti law (Calzetti 2001), respectively. The {\it GALEX} colours
predicted by the radiative transfer simulations are divided into several
bins in the IRX. The vertical error-bars show the bin widths. 
The filled circles and horizontal error-bars indicate the median
location and the full width of the distribution of the colours in each
bin. Six dust models are shown in each own panel from (a)--(f) as
indicated in the panels.

The SMC (WG) case (panel [b]) shows a very good agreement with the data
of the UV selected galaxies and the FIR selected galaxies with 
$F_{\rm dust}/F_{FUV}\lesssim100$. The LMC av (D) and the LMC 2
(D) cases (panels [e] and [f]) are also compatible with the data. The
colours predicted by the MW (D) case (panel [c]) are still
$\sim0.2$--0.3 mag bluer than the observed ones because of a strong bump
and a shallow UV slope in the extinction law, although a different IMF
like a Kroupa IMF could reduce the discrepancy (Panuzzo et al.~2006). 
On the other hand, the predicted colours of the MW (WG) case (panel [a])
are largely separated from the observed data, say $\sim0.5$ mag at  
$F_{\rm dust}/F_{FUV}\sim10$. For the SMC (D) case (panel [d]), the
predicted colours are too red ($\sim0.5$ mag at 
$F_{\rm dust}/F_{FUV}\sim10$) because of a rapid decrease of the albedo
between the two {\it GALEX} bands as shown in Fig.~1 (b).

There is no model which reproduces the FIR selected galaxies with 
$F_{\rm dust}/F_{FUV}\gtrsim100$. For an opaque disc, we expect to have
an attenuation law (i.e. transmission rate curve) independent of 
dust properties. Indeed, we find that the locations of the most opaque
point in each panel are very similar; all dust models predict a very
similar position on the diagram for $F_{\rm dust}/F_{FUV}\gtrsim100$. 
However, the real galaxies show a very large dispersion in the region. 
Burgarella et al.~(2005a) suggested that an effect of ``decoupling'' is
important for such galaxies; stellar populations producing the UV and
the IR are completely different. For example, the UV radiation comes
from the population outside the obscured region, whereas the population
heating dust which emits the IR radiation is embedded there. In this
case, the UV colour is decoupled with the UV attenuation traced by 
$F_{\rm dust}/F_{FUV}$.
In the framework of Inoue et al.~(2006), such a ``decoupling'' would
take place if we consider an intermittent SFH. Now, we have three stellar
populations: young and intermediate ones embedded in clumps and in the
dusty disc, and old one distributed diffusely to the outside of the
disc. Under an intermittent SFH with a time-scale longer than $\sim300$
Myr (age threshold between the intermediate and old populations), we can
expect that the luminosity weights strongly vary along the time, and
then, the position of the most opaque case on the IRX--UV colour diagram
would vary. 

In summary, there are two sorts of dust suitable for the {\it GALEX}
colours of nearby ``normal'' galaxies: (1) dust with a bump and a
smaller albedo for a shorter wavelength like the Draine's MW, LMC av,
and LMC 2 models, and (2) dust without any bump but an almost constant
albedo like the WG's SMC model. Neither dust with a bump and a constant
albedo (except for the bump range) like the WG's MW model, nor dust
without any bump and a smaller albedo for a shorter wavelength like the
Draine's SMC model are suitable for the nearby galaxies observed with
the {\it GALEX}. This competitive relation between the albedo and the
bump should be robust if we change the adopted dust models.

\section{Discussion}

We have found two types of dust suitable for the nearby galaxies: one
has a bump, the other does not have any bump. Here we discuss which type
is more favorable from a viewpoint of the bump carrier search.

A suggested candidate of the bump carrier is very small carbonaceous
grains like PAHs (Polycyclic Aromatic Hydrocarbons; L{\'e}ger et
al.~1989), QCCs (Quenched Carbonaceous Composites; Sakata et al.~1983), 
and UV processed HACs (Hydrogenerated Amorphous Carbon grains; 
Mennella et al.~1998), although this is not settled yet (Henning et
al.~2004). These very small carbonaceous particles are confidently
attributed to the unidentified infrared (UIR) emission band in 3--13
$\mu$m (L{\'e}ger \& Puget 1984, Sakata et al.~1984). The UIR emission
band is quite common in the ISM of the MW (e.g., Onaka 2004) and of
other galaxies (e.g., Genzel \& Cesarsky 2000), except for
low-metallicity ($\lesssim 1/5$ $Z/Z_\odot$) galaxies in which the UIR
emission is weak or absent (Engelbracht et al.~2005). If the very small
carbonaceous grains producing the UIR emission are really responsible
for the bump, we should find the bump in other galaxies (but not so
low-metallicity). Indeed, the bump found in M31 (Bianchi et al.~1996)
and some distant galaxies, for example, a lensing galaxy at $z=0.83$ 
(Motta et al.~2002) and Mg {\sc ii} absorption systems at $z=1.5$ 
(Wang et al.~2004). There are also signs of the bump imprinted in the
observed UV spectra of a galaxy at $z=0.048$ (Burgarella et al.~2005b),
of some {\it IUE} starburst galaxies (Noll \& Pierini 2005), 
and of some star-forming galaxies at $z\sim2$ (Noll \& Pierini 2005).
Therefore, dust with a bump is more favorable if the UIR carrier and the
bump carrier are the same, the very small carbonaceous grains.
For a firm conclusion, however, we would need more investigations.

\section*{Acknowledgments}

This paper is based on ``Effects of dust scattering albedo and 2175 \AA\
bump on ultraviolet colours of normal disc galaxies'', which has been
accepted to be published in the MNRAS. 
The author thanks co-authors of the paper,
Veronique Buat, Denis Burgarella, Pasquale Panuzzo, Tsutomu T.\
Takeuchi, and Jorge Iglesias-P{\'a}ramo for stimulating discussions 
very much.

\section*{References}

\begin{list}{}{%
 \setlength{\leftmargin}{0pt}
 \setlength{\parsep}{-5pt}
}%

\item
Bell, E. F. 2002, ApJ, 577, 150

\item
Bianchi, L., Clayton, G. C., Bohlin, R. C., et al. 1996, ApJ, 471, 203

\item
Buat, V., Iglesias-P{\'a}ramo, J., Seibert, M., et al. 2005, ApJ, 619, L51

\item
Burgarella, D., Buat, V., Iglesias-P{\'a}ramo, J. 2005a, MNRAS, 360, 1413

\item
Burgarella, D., Buat, V., Small, T., et al. 2005b, ApJ, 619, L63

\item
Calzetti, D. 2001, PASP, 113, 1449

\item
Calzetti, D., Kinney, A. L., Storchi-Bergmann, T. 1994, ApJ, 429, 582

\item
Draine, B. T. 2003, ApJ, 598, 1017

\item
Engelbracht, C. W., Gordon, K. D., Rieke, G. H., et al. 2005, ApJ, 628, L29

\item
Genzel, R., Cesarsky, C. J. 2000, ARA\&A, 38, 761

\item
Gordon, K. D. 2004, Astrophysics of Dust, ASP Conference Series, 309, 77

\item
Granato, G. L., Lacey, C. G., Silva, L., et al. 2000, ApJ, 542, 710

\item
Henning, Th., J{\"a}ger, C., Mutschke, H. 2004, 
Astrophysics of Dust, ASP Conference Series, 309, 603

\item
Iglesias-P{\'a}ramo, J., Buat, V., Takeuchi, T. T., et al. 2006, ApJS,
in press (astro-ph/0601235)

\item
Inoue, A. K. 2005, MNRAS, 359, 171

\item
Inoue, A. K., Buat, V., Burgarella, D., et al. 2006, MNRAS, in press 
(astro-ph/0605182)

\item
Kong, X., Charlot, S., Brinchmann, J., et al. 2004, MNRAS, 340, 769

\item
L{\'e}ger, A., Puget, J.-L. 1984, A\&A, 137, L5

\item
L{\'e}ger, A., Verstraete, L., D'Hendecourt, L., et al. 1989, 
Proceedings of IAU symposium, 135, 173

\item
Mannella, V., Colangeli, L., Bussoletti, E., et al. 1998, ApJ, 507, L177

\item
Meurer, G. R., Heckman, T. M., Calzetti, D. 1999, ApJ, 521, 64

\item
Noll, S., Pierini, D. 2005, A\&A, 444, 137

\item
Onaka, T. 2004, Astrophysics of Dust, ASP Conference Series, 309, 163

\item
Panuzzo, P., Granato, G. L., Buat, V., et al. 2006, MNRAS, submitted

\item
Robin, A. C., Reyl{\'e}, C., Derri{\`e}re, S., et al. 2003, A\&A, 409, 523

\item
Sakata, A., Wada, S.,  Okutsu, Y., et al. 1983, Nature, 301, 493

\item
Sakata, A., Wada, S., Tanab{\'e}, T., et al. 1984, ApJ, 287, L51

\item
Seibert, M., Martin, C. D., Heckman, T. M., et al. 2005, ApJ, 619, L55

\item
V{\'a}rosi, F., Dwek, E. 1999, ApJ, 523, 265

\item
Wang, J., Hall, P. B., Ge, J., et al. 2004, ApJ, 609, 589

\item
Weingartner, J. C., Draine, B. T. 2001, ApJ, 548, 296

\item
Witt, A. N., Gordon K. D. 2000, ApJ, 528, 799

\item
Wolfire, M., G., McKee, C. F., Hollenbach, D., et al. 2003, ApJ, 587, 278
\end{list}

\end{document}